\documentclass[twocolumn]{aastex631}
\usepackage{graphicx}

\received{June 24, 2021}
\accepted{September 15, 2021}

\shorttitle{Are Milky Way dwarfs at first infall?}
\shortauthors{Hammer et al.}

\begin{document}

\title{Gaia EDR3 proper motions of Milky Way dwarfs. II: Velocities, Total Energy and Angular Momentum}

\correspondingauthor{Francois Hammer}
\email{francois.hammer@obspm.fr}

\author[0000-0002-2165-5044]{Francois Hammer}
\affiliation{GEPI, Observatoire de Paris, Universit\'e PSL, CNRS, Place Jules Janssen F-92195, Meudon, France}
\author{Jianling Wang}
\affiliation{CAS Key Laboratory of Optical Astronomy, National Astronomical Observatories, Beijing 100101, China}
\author[0000-0002-9197-9300]{Marcel S. Pawlowski}
\affiliation{Leibniz-Institut fuer Astrophysik Potsdam (AIP), An der Sternwarte 16, D-14482 Potsdam Germany}
\author{Yanbin Yang}
\affiliation{GEPI, Observatoire de Paris, Universit\'e PSL, CNRS, Place Jules Janssen F-92195, Meudon, France}
\author[0000-0002-1014-0635]{Piercarlo Bonifacio}
\affiliation{GEPI, Observatoire de Paris, Universit\'e PSL, CNRS, Place Jules Janssen F-92195, Meudon, France}
\author[0000-0002-9497-8127]{Hefan Li}
\affiliation{School of Physical Sciences, University of Chinese Academy of Sciences, Beijing 100049, P. R. China}
\author[0000-0002-7631-348X]{Carine Babusiaux}
\affiliation{Universit\'e de Grenoble-Alpes, CNRS, IPAG, F-38000 Grenoble, France }
\affiliation{GEPI, Observatoire de Paris, Universit\'e PSL, CNRS, Place Jules Janssen F-92195, Meudon, France}
\author[0000-0003-2837-3899]{Frederic Arenou}
\affiliation{GEPI, Observatoire de Paris, Universit\'e PSL, CNRS, Place Jules Janssen F-92195, Meudon, France}



\begin{abstract}
 Here we show that precise Gaia EDR3 proper motions have provided robust estimates of 3D velocities, angular momentum, and total energy for 40 Milky Way dwarfs. The results are statistically robust and are independent of the Milky Way mass profile. Dwarfs do not behave like long-lived satellites of the Milky Way because of their excessively large velocities, angular momenta, and total energies.  Comparing them to other MW halo populations, we find that many are at first passage, $\le$ 2 Gyr ago, i.e., more recent than the  passage of Sagittarius, $\sim$ 4-5 Gyr ago. We suggest that this is in agreement with the stellar populations of all dwarfs, for which we find that a small fraction of young stars cannot be excluded. We also find that dwarf radial velocities contribute too little to their kinetic energy when compared to satellite systems with motions only regulated by gravity, and some other mechanism must be at work such as ram pressure. The latter may have preferentially reduced radial velocities when dwarf progenitors entered the halo until they lost their gas. It could also explain why most dwarfs lie near their pericenter.  We also discover a novel large-scale structure perpendicular to the Milky Way disk, which is made by 20\% of dwarfs orbiting or counter-orbiting with the Sagittarius dwarf. 

\end{abstract}



\section{Introduction} \label{sec:intro}
Gaia EDR3 has provided robust orbital parameters for 46 dwarfs (classical dwarf spheroidal, dSphs, and ultra-faint dwarfs) in the Milky Way (MW) halo, with a precision that is on average 2.5 times better than that from Gaia DR2  \citep[Paper I]{Li2021}. \citet{Li2021} also showed that the major uncertainty in establishing integrated orbital properties  (apocenters, orbit shapes, and eccentricities) is due to our lack of knowledge of the MW potential, though intriguingly the same does not apply for the pericenter determination. By comparing their Tables 4 and 5 one finds that the fraction of unbound dwarfs (defined with $P_{\rm unbound} >$ 50\%) decreased from 56\% to 13\% when passing from a low mass ($M_{\rm tot}$= 2.8 $10^{11}M_{\odot}$) to a five times more massive  MW model ($M_{\rm tot}$= 15 $10^{11}M_{\odot}$), covering the full mass range able to reproduce the rotation curve (see  \citealt{Jiao2021}). \\

Such an uncertainty in orbital properties hampers our ability to evaluate the nature and origin of dwarfs. Could one consider MW dwarfs as bound satellites and then derive the MW mass accordingly \citep{Callingham2019,Cautun2020}? Are MW dwarfs long-lived satellites of the MW, or are most of them at first passage such as the Magellanic Clouds?  \citet{Kallivayalil2013} suspected the latter to be at first approach on the basis of the MW mass but also on that of the LMC mass, which if supposed to be very massive ($\geqslant 10^{11}$ $M_{\odot}$) could have its orbital motion decayed through, e.g., dynamical friction. \citet{Kallivayalil2013} based their analyses upon the tidal model of \cite{Besla2012}, which supposed by design that the SMC has been gravitationally bound to the LMC for a few billion years. However the Magellanic stream properties, such as its double filamentary structure, are apparently better reproduced by a ``ram pressure + collision'' model \citep{Hammer2015, Wang2019}, which leads to a much less massive LMC, with a mass smaller than $2\times10^{10}$ $M_{\odot}$. Nevertheless, the uncertainty linked to the MW mass is truly limiting our knowledge of the nature of its companions. \\

Star formation histories (SFHs) of dwarfs are often considered as a major argument in favor of the long-lived satellite scenario, for which they have progressively fallen into the halo within a Hubble time, in agreement with expectations for halos in the $\Lambda$CDM paradigm. SFHs of classical dwarfs can be either dominated by a single, very old population with no star formation in the last eight billion years (Canes Venaciti I, Sculptor, and Draco), or can be extended to the two last billion years (Carina, Fornax, Leo I, and LeoII) as shown by \cite[see also \citealt{Hurley-Keller1998} for Carina]{Weisz2014}, and by \citet[see also \citealt{Komiyama2007} for Leo II]{Zhang2017}. However, global SFH information might be superseded by more-in-depth studies showing that classical dSphs can be described by the superposition of two or more populations \citep[and references therein]{Pace2020}, showing a dynamically colder component with higher metallicity or smaller age \citep[see also dynamical differences in the two Sculptor populations found by \citealt{Zhu2016}]{Komiyama2007} in the center when compared to the outskirts. Whether or not dwarfs are fully devoid of moderately young, $\le$ 2 Gyr old stars is still an open question. The situation is even less clear for the fainter dwarfs because they are generally lacking an appropriate number of RGB stars to sufficiently populate their color-magnitude diagrams (see the discussion in \citealt{Brown2014}).\\

Many physical processes may affect the dSph SFHs, which deserve a careful analysis to constrain their infall time. Before their infall, dSph progenitors were likely gas-rich dwarf irregulars \citep{Grcevich2009,Hammer2018a}. Ram pressure due to the MW halo gas may remove the gas during the infall, and it may likely induce the last star formation episode, depending on the precise orbit. Such an event has been successfully identified in good agreement with the Leo I orbital motion, and has been interpreted to occur at the pericenter passage \citep{Ruiz-Lara2021}. Fornax has a recent star formation \citep{Weisz2014}, and if we were observing this galaxy 50-100 Myr ago, it would have looked like a dwarf irregular \citep{Battaglia2012}. Fornax could be an archetype for a first infall recently affected by ram pressure (see Yang et al. 2021, submitted to MNRAS). It is not necessary that a large fraction of stars formed during that episode. For example, the LMC has formed only 15\% of its stellar mass within the last two billion years \citep{Harris2009}, and it is still affected by ram-pressure that likely induces its most active star-forming region (30 Doradus).\\

If SFHs are insufficient or dubious for inferring the dwarf past orbital history, could the later be directly constrained by the 6D space-velocity coordinates, given the precision derived for many of them by Gaia EDR3?  This paper aims at investigating this question. Here, we prioritize  the study of instantaneous quantities to preserve our independence of the unknown MW potential. Alternatively, each MW-mass-dependent quantity will be evaluated within the whole range of MW masses consistent with the MW rotation curves \citep{Eilers2019,Mroz2019}. \\

In Section 2 we will evaluate whether dwarfs show excessively large tangential velocities, when compared to expectations from cosmological simulations. In Section 3, we will trace the former orbital history of the MW halo, by comparing dSph locations in the energy-angular momentum plane with those of globular clusters, Sagittarius stream stars, and the old population of K-giant stars. In Section 4, we tentatively identify a novel large-scale structure linked to Sagittarius, and we discuss in Section 5 whether MW dwarfs are satellites or are, alternatively, at first passage.

\section{Are MW dwarfs consistent with a long-lived satellite system?} 
\subsection{The excess of dSph tangential velocities}
\label{sec:beta}
A number of previous studies have investigated the velocity anisotropy of the MW dwarfs when they are considered as being MW satellites. Here we review results based on different dwarf samples and recalculate the anisotropy for each sample using the EDR3 data (see also Table~\ref{tab:aniso} for a summary). 

\citet{Cautun2017} examined the tangential velocity excess of MW dwarfs and concluded that such an excess can be reproduced by only a few percent of $\Lambda$CDM cosmological realizations of satellite systems. To characterize the velocity anisotropy they have used:
\begin{equation}
\beta= 1 - \Sigma (V_{tan}^2)/(2 \times \Sigma(V_{rad}^2))
\label{eq:beta}
\end{equation}

for which the sum has been made for the eight classical dwarfs (Carina, Draco, Fornax, Leo I, Leo II, Sagittarius, Sculptor, and Ursa Minor), plus the Magellanic Clouds. Based on quantities derived from HST-based proper motions, \citet{Cautun2017} found $\beta$= -2.2 instead of 0.5 that is the most expected value for $\Lambda$CDM simulated subhalos, whose motions are dominated by radial velocities. Using Gaia EDR3 we confirm this discrepancy, with  $\beta$= -1.36 for the \citet{Cautun2017} sample of the 10 most luminous dwarfs, a value very close to that (-1.5) found by \citet{Riley2019}. \\

The latter study encompassed the Gaia DR2 results from \citet{Fritz2018}, gathering 36 dwarfs plus the Magellanic Clouds. They found that $\beta$ takes negative values for dwarfs within a 100 kpc distance ($\beta$ $\sim$ -1) but becomes positive for dwarfs more distant than 100 kpc ($\beta$ $\sim$ 0.3, see their Fig. 5). \citet{Riley2019} concluded on the absence of a tangential velocity excess for distant MW dwarfs, and suggested that dwarfs in the inner halo could be affected by the MW disk. The latter would have preferentially destroyed satellites on radial orbits, explaining then the $\beta$ radial profile.\\

Gaia EDR3 provides discrepant results from \citet{Riley2019} for the subsample of dwarfs at distances larger than 100 kpc. Using 46 dwarfs from \citet{Li2021}, we find $\beta$= -1.03$\pm$0.15 (-0.95) at $r_{\rm GC}<$ 100 kpc and $\beta$= -2.6$\pm$1.3 (-0.91) at $r_{\rm GC}>$ 100 kpc, where the values in parentheses are coming from using the same subsample of \citet{Riley2019}, dubbed the 'gold  sample'. 
To further verify whether our results are consistent within the \citet{Riley2019} formalism, we have recalculated $\beta$ using their Equation (1)\footnote{$\beta= 1 - (\sigma_{\theta}^2+\sigma_{\phi}^2)/(2 \times \sigma_{rad}^2)$, where $\sigma_{\theta}$, $\sigma_{\phi}$, and $\sigma_{\rm rad}$ are the velocity dispersions along each coordinate direction, and tabulated in Table 2 of \citet{Li2021}.} instead of Eq.~\ref{eq:beta} above, which is based on \citet{Cautun2017}. This leads to almost unchanged values of $\beta$= -0.8 and -2.0 for the nearby ($r_{\rm GC}<$ 100) and a distant ($r_{\rm GC}>$ 100 kpc) subsample, respectively. It is possible that the discrepancy is caused by the lack of precision from Gaia DR2, which affects especially the tangential velocity values of distant dwarfs. 

\begin{figure}
\includegraphics[width=3.4in]{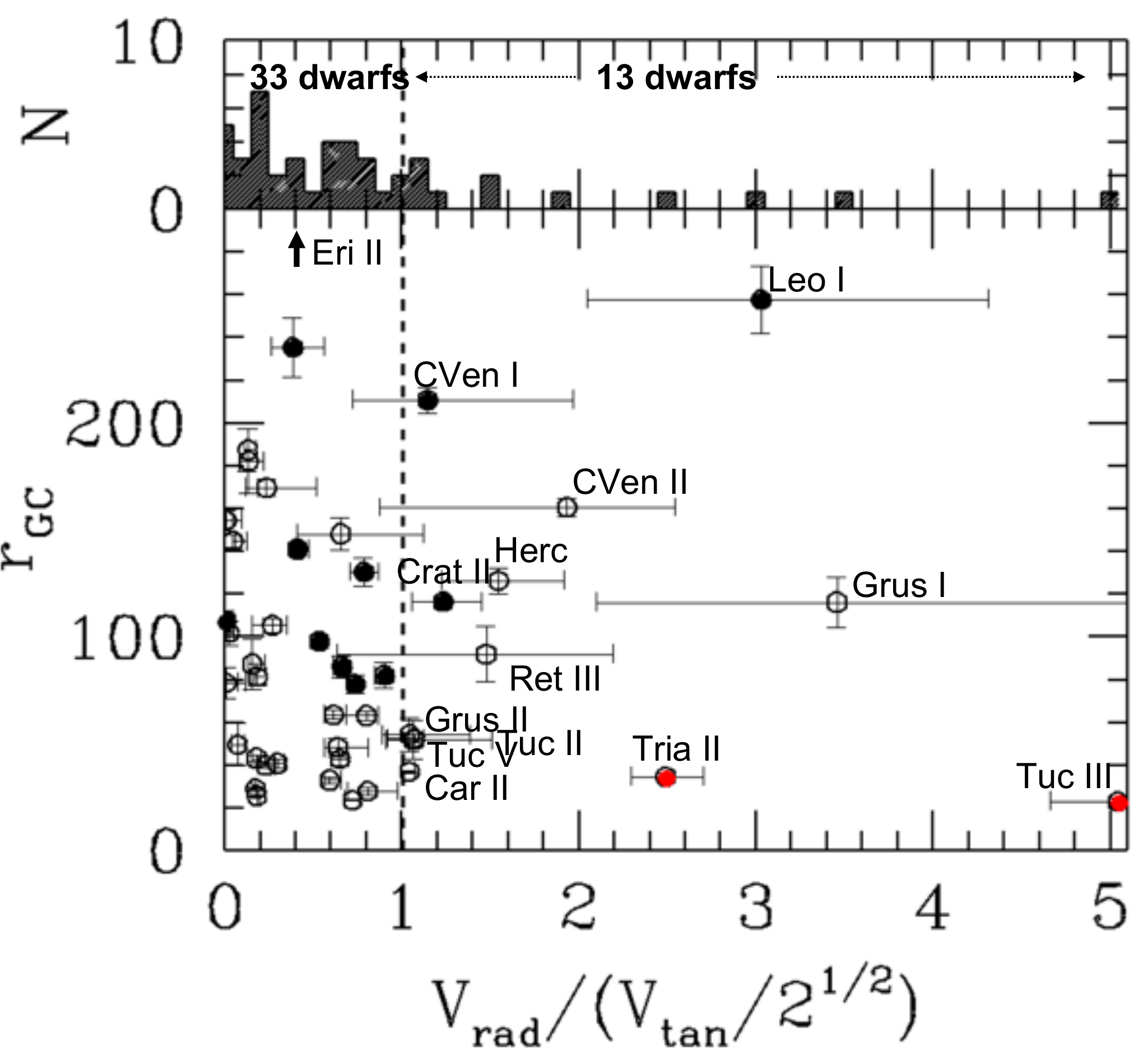}
\caption{Top:  histogram of the $V_{rad}/(V_{tan}/\sqrt{2})$ ratio, which should be centered around 1 (see the vertical dotted line) if 3D velocities were spatially distributed at random. Number of dwarfs are indicated below and above 1. Bottom: the $V_{rad}/(V_{tan}/\sqrt{2})$ ratio is represented against the distance to the Galactic center, $r_{\rm GC}$. Eridanus II at a distance of 366 kpc is not represented here (see the vertical arrow) and its ratio is 0.4$\pm$0.25. The 13 objects with an excess of radial velocities ($V_{rad}/(V_{tan}/\sqrt{2}) >1$) are labeled and Tucana III and Triangulum II are distinguished by a red dot since their pericenters are smaller than 20 kpc (see the text).}
\label{fig:rp_VradvsVtan}
\end{figure}

 This is illustrated in the top panel of Figure~\ref{fig:rp_VradvsVtan} showing that 33 dwarfs have $V_{rad}/(V_{tan}/\sqrt{2})$$<$ 1 corresponding to $\beta < 0$, which could be compared to the 13 (labeled) dwarfs with $V_{rad}/(V_{tan}/\sqrt{2})$$>$ 1. We have further investigated the \citet{Riley2019}'s conjecture that satellites passing on radial orbits near the MW disk are preferentially destroyed leading to a bias in favor of high tangential velocities at small pericenter. There are two objects\footnote{The impact of the MW disk likely decreases at pericenters reaching 20-25 kpc (Draco II, Grus I, and Segue I; 3 dwarfs that do not show excessive tangential velocities - the radial component even dominates for Grus I).} with pericenters sufficiently small to be well inside the disk, Tucana III ($r_p$= 3 kpc) and Triangulum II ($r_p$= 12 kpc), and both show radial velocities (-227.6 and -259.8 km $s^{-1}$) much larger than their tangential velocities (65.1 and 146.4km $s^{-1}$). Both are plunging into the MW on a destructive orbit, confirmed by the fact that Tucana III is embedded into a stellar stream \citep{Mutlu-Pakdil2018}, a feature also suspected for Triangulum II  \citep{Martin2016}. However, it is difficult to verify the impact of the disk in destroying dwarfs with too many radial orbits, since by construction, these putative dwarfs have disappeared from the dwarf sample. \citet{Riley2019} made this prediction to explain the absence of velocity anisotropy at distances larger than 100 kpc, which is not further verified with Gaia EDR3. If correct, there should be more tangentially dominated velocities (with $V_{rad}/(V_{tan}/\sqrt{2})$$<$ 1) at low radii than at large radii. Comparing the two intervals, one with 100 $<$ $r_{\rm GC}$ $<$ 200 kpc (and the other with 20 $<$ $r_{\rm GC}$ $<$ 100 kpc), one finds 11 (20) dwarfs with $V_{rad}/(V_{tan}/\sqrt{2})$$<$ 1 among 15 (27), respectively, i.e., $\sim$ 75\% in both distance intervals. Based on these numbers, the conjecture of Riley et al. (2019) does not appear to have an obvious signature in the present dataset.

\begin{deluxetable}{LCCC}
\label{tab:aniso}
\tablecaption{Velocity anisotropy $\beta$ according to different studies}
\tablehead{
	\colhead{Samples/studies} & \colhead{A} & \colhead{B} & \colhead{C} 
}
\decimals
\startdata
10luminous-dwarfs & -2.2 &  -1.5 & -1.36 \\
Gaia DR2/EDR3  & - &  -1.02$\pm$0.4 & -1.47$\pm$0.47\\
$\le$100 kpc   & - & -1  & -1.03$\pm$0.15  \\
$>$100 kpc & -  &  0.3& -2.6$\pm$1.3 \\
\enddata
\tablecomments{Studies are A: \citet{Cautun2017}, B: \citet{Riley2019}, Gaia DR2, 36 dwarfs, and C: this study, EDR3, 46 dwarfs}
\end{deluxetable}

 \subsection{Could circular motions explain high $V_{\rm tan}$ and dwarf proximity to their pericenters?}
\label{sec:peri}
Figure~\ref{fig:rp_VradvsVtan} reveals that 33 (72\%) dwarfs among 46 have excessive tangential velocities ($V_{rad}/(V_{tan}/\sqrt{2})$$<$ 1).  In the long-lived satellite paradigm, excessive tangential velocities ( $\beta$ $<<$ 0, or $V_{rad}/(V_{tan}/\sqrt{2})$$<<$ 1, see the right panel of Figure~\ref{fig:rp_VradvsVtan})   would be simply explained if dwarf orbits were circular \citep{Cautun2017,Riley2019}. This would also have the advantage of reconciling the long-lived satellite paradigm with the fact that many dwarfs lie near their pericenters \citep{Fritz2018,Simon2018,Hammer2020,Li2021}. To test this conjecture, we have examined the eccentricities of dwarfs with large tangential velocities, and compared them with the rest of the sample. Furthermore, we have only considered MW models with large masses from \citet[the two models with total masses of $15\times10^{11}$ and of $8.1\times10^{11}$  $M_{\odot}$]{Li2021} for which most dwarfs are bound. \\

We have first identified the 21 dwarfs for which $V_{\rm tan}$ values are so large that they fully dominate the kinetic energy (equivalent to $\beta$$<$-2 or $V_{rad}/(V_{tan}/\sqrt{2})$$<$ 0.57). This subsample is ideal to test whether or not their orbits are more circular than for the rest of the sample. The 21 dwarfs with $V_{\rm 3D}$ $\sim$ $V_{\rm tan}$ do not have circular orbits and their average eccentricity is 0.57 (0.80), which is identical to 0.6 (0.75) for the whole sample for the largest (second largest) MW mass from \citet{Li2021}, respectively. Excess of tangential velocities for many dwarfs cannot be explained by circular orbits, and it still challenges  their interpretation as being long-lived satellites of the MW. \\

\begin{figure*}
\includegraphics[width=6.8in]{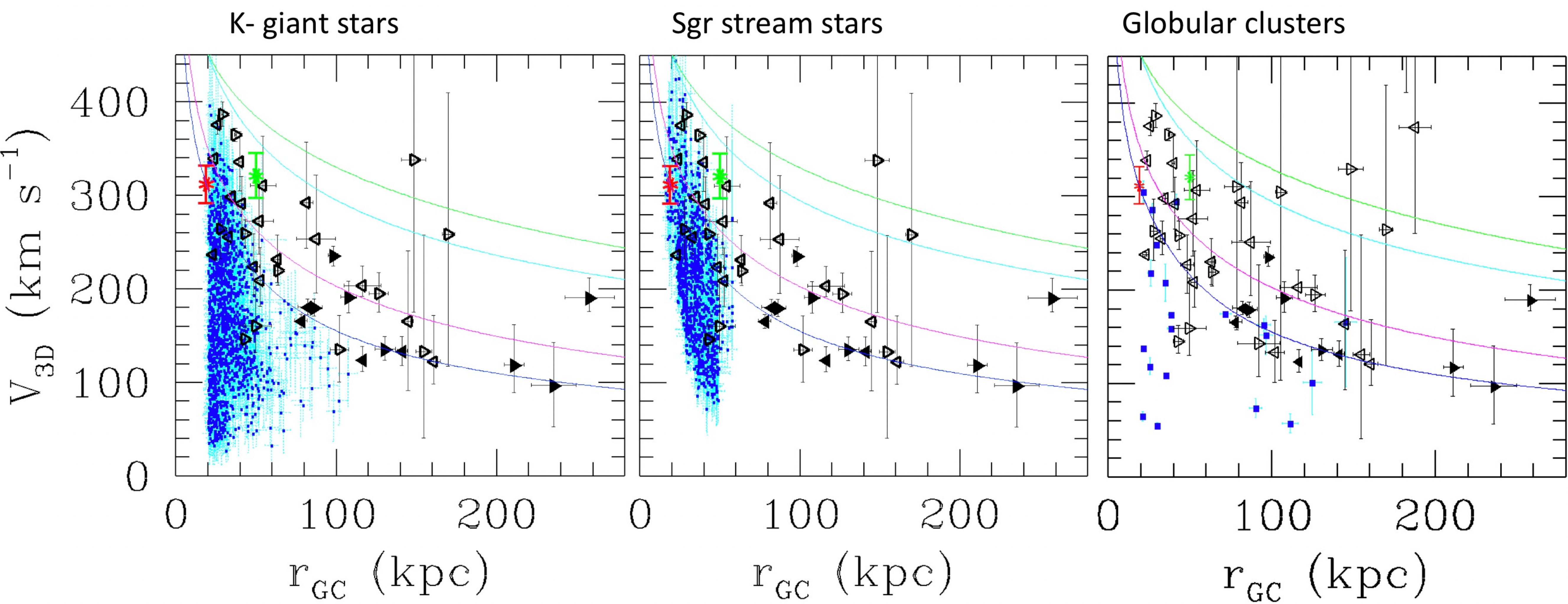}
\caption{Phase diagram for 40 dwarfs (triangles) defined with robust proper motions in \citet{Li2021}, i.e., excluding Aquarius II, Columba I, Horologium II, Pisces II, Reticulum III (fewer than three stars for proper motions), and Eridanus II. Open and full triangles are for stellar masses smaller or larger than $10^{5}$ $M_{\odot}$, respectively. The  triangle orientation towards $r_\mathrm{GC}=0$ or in the opposite direction indicates whether dwarfs are on an approaching ($V_{\rm rad} < 0$) or on a receding ($V_{\rm rad} > 0$) orbit, respectively. Star points indicate Sagittarius (red) and the LMC (green) positions. Left:  this panel shows the comparison with old K-giant stars (EDR3, blue squares, cyan error bars).  Middle: same, but with comparison to Sagittarius stream RGB stars (DR2; from \citealt{Vasiliev2021b}). Right: dwarfs (triangles) compared to globular clusters (blue squares, EDR3) from \citet{Vasiliev2021a}. Each dSph is labeled by its short name. Green, cyan, magenta, and blue lines indicate the escape velocity of the four MW mass models adopted in \citet{Li2021}: Einasto high-mass, PNFW (from \citealt{Eilers2019}), intermediate mass, and low mass, with total MW masses of $15\times10^{11}$, $8.1\times10^{11}$, $5.1\times10^{11}$, and  $2.8\times10^{11}$ $M_{\odot}$, respectively.}
\label{fig:phase}
\end{figure*}

\section{Energy and angular momentum : infall history of the halo populations} 
\label{sec:energy}

In this section, we aim at comparing the different populations inhabiting the MW halo, and in particular their energy and angular momentum, which are fundamental quantities resulting from their past orbital histories. Since we have several constraints on the orbital history of K-giant stars, Sagittarius stream stars, and globular clusters, the comparison may help us to constraint the dwarf orbital history.  As our dwarf samples all have r $>$ 20 kpc, we similarly limit star and GC samples to r $>$ 20 kpc for consistency. The impact of a possible massive LMC would require further investigation to see how it may affect the kinematics of the various tracers considered here, which is beyond the scope of this paper (but see Wang et al. 2021, submitted to MNRAS).

\subsection{Gaia EDR3 proper motions of halo stars and globular clusters}
The K giant stars are selected from the SDSS/SEGUE photometry and spectroscopy survey
\citep{Xue2014,Xue2015},
which provide distances and sky positions.
We cross-match both the K giant star catalog with Gaia EDR3 by 1 arc-second
tolerance to obtain proper motion and their errors.
Sgr stream star members are excluded following the angular momentum method of
\citet{Petersen2021}.\\

RGB stars from the Sagittarius dSph and its stream are coming from \citet{Vasiliev2021b}, who cross-matched the Gaia DR2 source catalog with the 2MASS catalog
to select samples. This is further complemented by large spectroscopic surveys to add radial velocities using $APOGEE$, $LAMOST$, and $SDSS$. \citet{Vasiliev2021b} made a careful selection based on  various criteria to select a clean sample devoid of halo stars of different origin. \\

The globular clusters are taken from Wang et al. (2021, submitted to MNRAS), who have measured the mean PM and its associated errors with Gaia EDR3 following the method of \citet{Vasiliev2019}.  Wang et al. (2021, submitted to MNRAS) have followed the exact method of \citet{Vasiliev2021b}, and  have used a probabilistic gaussian mixture model to determine the membership probability, and the error-deconvolve intrinsic parameters for both member stars and non-member stars. We further note that there are only 20 globular clusters with $r_{\rm GC} > $ 20 kpc.

\subsection{Comparison to other populations of the MW halo}

Figure~\ref{fig:phase} compares the phase diagrams of 40 dwarfs with robust proper motions \citep{Li2021} to that of other MW halo populations, namely the old K-giant stars, the Sagittarius stream stars, and the globular clusters. It indicates three main global features:
\begin{itemize}
\item Observed stars and globular clusters occupy a much more restricted space in the halo than dwarfs.
\item In the volume they are coinhabiting (from 20 to about 60 kpc) old K-giant stars, and to a lesser extent, Sagittarius stream stars have smaller 3D velocity (average $V_{\rm 3D}$= 174 and 209 km$s^{-1}$, respectively) or kinetic energy than dwarfs (average $V_{\rm 3D}$= 279 km$s^{-1}$).
\item  Globular clusters systematically show smaller 3D velocities (average $V_{\rm 3D}$= 164 km$s^{-1}$)  than dwarfs  (average $V_{\rm 3D}$= 241 km$s^{-1}$) from $r_{\rm GC} $ = 20-140 kpc.
\end{itemize}

\begin{figure*}
\includegraphics[width=6.8in]{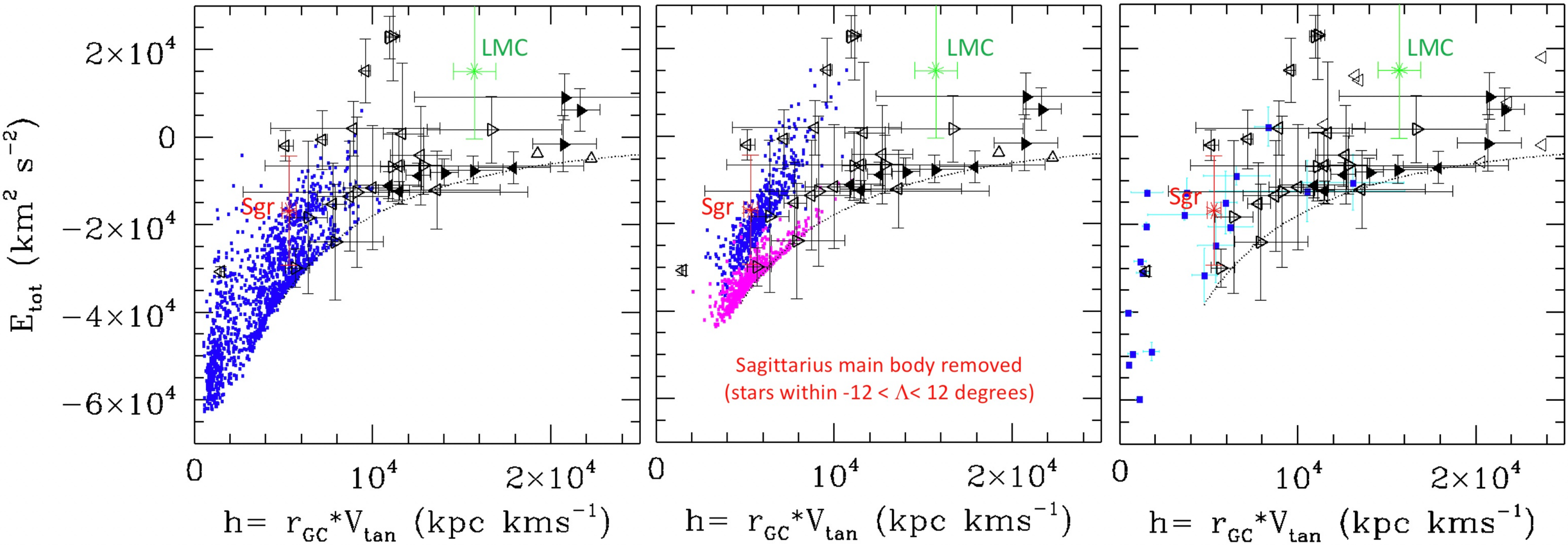}
\caption{
Total energy vs. angular momentum for 30 dwarfs, and assuming the Milky-Way intermediate mass model of \citealt[see also Wang et al. 2021]{Li2021}. Dwarfs are shown by triangles as in Figure~\ref{fig:phase}, and have been selected with uncertainties $\Delta h$ $<$ $10^{4}$ kpc $\times$ km $s^{-1}$ and $\Delta E_{tot}$ $<$ 2 $10^{4}$ kpc $\times$ km $s^{-1}$ on energy). LMC and Sagittarius are represented by green and red stars, respectively. The dotted curve represents the energy-angular momentum relation for a particle of negligible mass in a circular orbit at virial equilibrium. From left to right: comparison with the old, halo K-giant stars, with the Sagittarius stream stars (magenta dots: Leading Arm and blue dots: Trailing Arm), and with globular clusters.}
\label{fig:E_AM_allpop}
\end{figure*}

To further investigate, we examine the angular momentum (h= $r_{\rm GC}$ $\times$ $V_{\rm tan}$) and total energy ($E_{\rm tot}$= 0.5 $\times$ $V_{3D}^2$ + $\Phi$),  plane, where $\Phi$ is the MW potential energy. If no energy is exchanged with the MW, both quantities are conserved, and they could reveal the past orbital histories of the MW halo populations. Figure~\ref{fig:E_AM_allpop} compares the dSph location in the ($h$, $E_{\rm tot}$) plane with that of old K-giant stars, Sagittarius stream stars, and globular clusters, respectively. The figure only shows dwarfs with errors on $h$ smaller than $10^{4}$ kpc $\times$ km $s^{-1}$, and errors on total energy smaller than 2 $10^{4}$ kpc $\times$ km $s^{-1}$, respectively\footnote{The reason to remove the 11 dwarfs with the largest errors is that their error bars are almost half as large as the amplitude of either $E_{tot}$ or $h$. When compared to Figure~\ref{fig:phase} this has removed some of the most distant dwarfs, Leo II and Canes Venaciti I, and also dwarfs with significant uncertainties on their $V_{\rm 3D}$ or $V_{\rm tan}$ values, Bootes II, Crater, Horologium I, Hydra II, Leo IV and V, Phoenix II, and Tucana II and V  (see Appendix~\ref{sec:h_E_err}). Besides this, Reticulum III, which was excluded in Figure~\ref{fig:phase} has been reintroduced because its error bars match our criteria.}. Old K-giant stars, Sagittarius stream stars, globular clusters, and dwarfs show a correlation between the total energy and the angular momentum. Dwarf galaxies show a considerable excess in both total energy and angular momentum when they are compared, in order of increasing differences, to Sagittarius stream stars, globular clusters, and old K-giant stars, respectively. In Figure~\ref{fig:E_AM_allpop} we have chosen the intermediate mass model of \citet{Li2021} for the MW\footnote{Note that a few K-giants and some Sgr stream stars seem to be unbound using this model. Accounting for error bars, we find that all K-giant stars are consistent with being bound, while $\sim$ 15 Sgr trailing stream stars appear to be unbound. Only a full model of the Sgr trailing arm may help to verify whether there could be unbound stars ejected by the MW tidal forces.} , which is also the one favored by Wang et al. (2021). Adopting another model would not change the above conclusions, because a change of the MW potential affects similarly each population inhabiting the halo (see Appendix~\ref{sec:MWmasses}).\\

\begin{figure}
\includegraphics[width=3.4in]{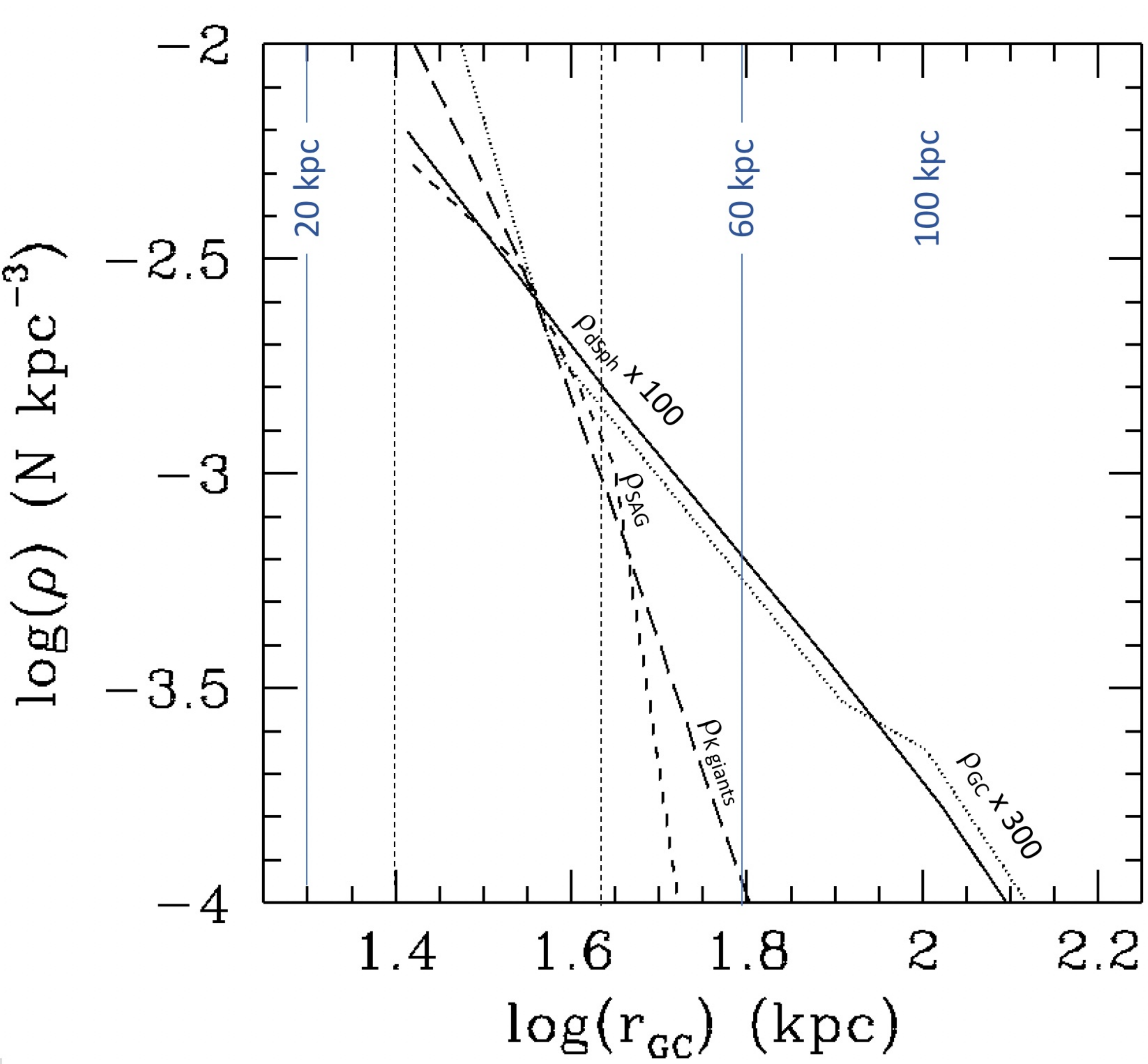}
\caption{ Number density radial profile of the K-giant stars (long dashed line), the Sgr stream stars (short dashed line), the dwarfs ($\times$ 100, solid line), and of globular clusters ($\times$ 300, dotted line). The two full vertical lines indicate the 20 and 60 kpc limit  imposed to compare the different samples in the left panel. The two vertical dotted lines represent the average radii in which the densities of stars have been recalibrated to match the density function of dwarf galaxies. }
\label{fig:Ndensities}
\end{figure}

\begin{figure*}
\includegraphics[width=6.5in]{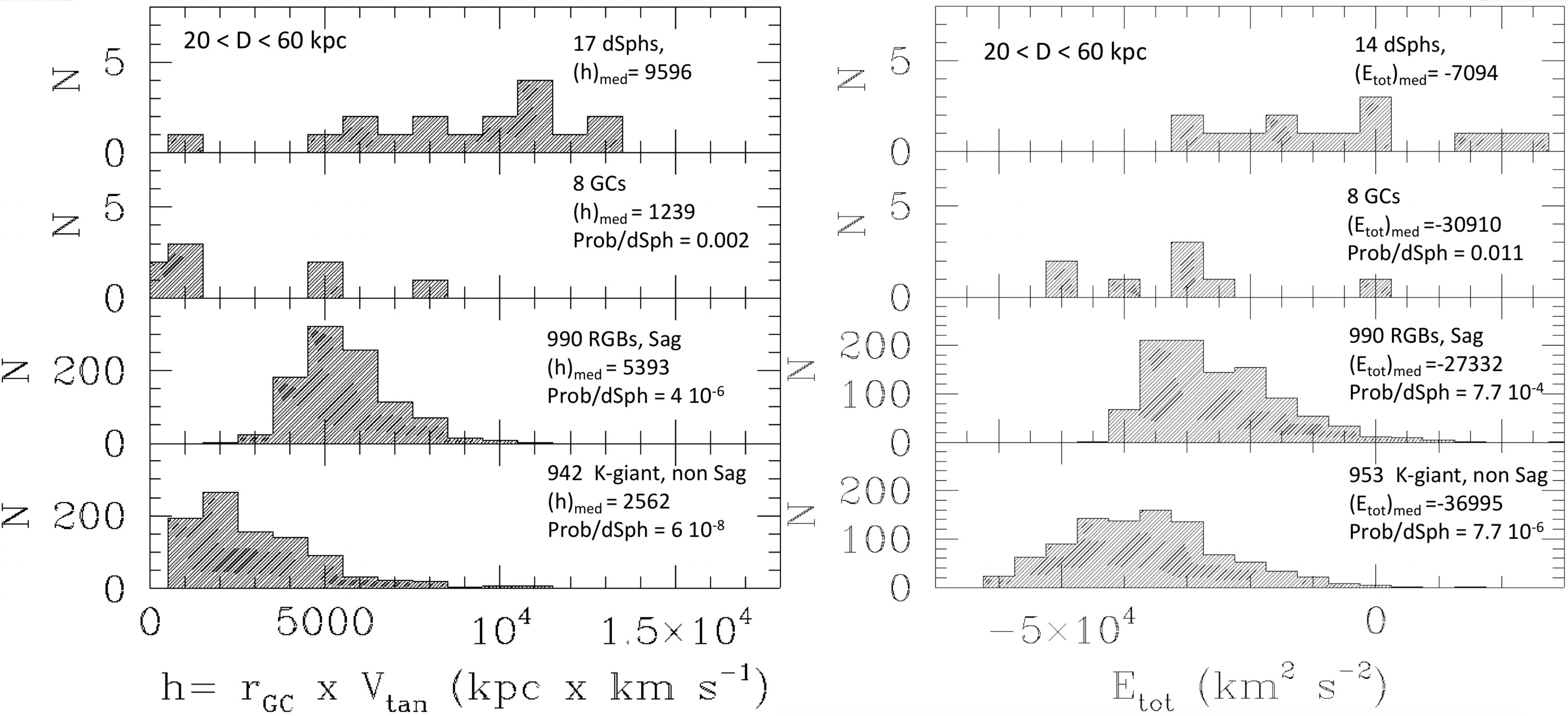}
\caption{Left: histograms of angular momentum for dwarfs (top), globular clusters (second row), Sgr stream stars (third row), and K-giant stars (bottom), all being selected between 20 and 60kpc. Number of objects for each species is written in each panel, and for stars has been statistically corrected to adjust the radial distribution function of dwarfs (see the text). The  median value of $h$ and the probability that the distribution is consistent with that of dwarfs are also indicated. Right: same as the left panel, after replacing $h$ by $E_{\rm tot}$.  Only 14 dwarfs, instead of 17, are represented, because here we have excluded dwarfs with large error bars in $E_{\rm tot}$ (see the text).}
\label{fig:histos}
\end{figure*}

Figure~\ref{fig:Ndensities} compares the radial profile of the density function for dwarfs, globular clusters, K-giant stars, and Sgr stream stars. As shown in Figure~\ref{fig:phase} dwarfs are associated with a more extended radial profile than other halo populations. To compare their angular momentum distribution requires to correct the distribution function of stars, in order to keep them consistent with that of dwarfs. First, we choose to compare all populations within an interval from 20 to 60 kpc. Second, we have calculated densities in two subintervals namely 20-30 kpc and 30-60 kpc, respectively. We have then removed, using a random selection, 70\% (25\%) of old K-giant (Sgr stream) stars at 20-30 kpc to match the dSph distribution function.  This has been done in the left panel of Figure~\ref{fig:histos}, which compares the different distributions of the different halo populations on the basis of a similar distribution function.\\

Given their small number, the comparison with globular clusters is not very meaningful, though they seem to systematically show a smaller angular momentum than dwarf galaxies. Besides this, the angular momentum of the latter are so large that their distribution is not consistent with that of K-giant stars (with a Kolmogorov-Smirnov probability that they come from the same distribution, P=6 $10^{-8}$), or with Sgr stream stars (P=4 $10^{-6}$), respectively.\\

The right panel of Figure~\ref{fig:histos} compares the total energy distribution of dwarfs, globular clusters, and K-giant and Sgr stream stars, as it has been done for the angular momentum (left panel).  The dwarfs'large values for the total energy distribution are not consistent with that of K-giant (with a probability P=7.7 $10^{-6}$) or Sgr stream (P=7.7 $10^{-4}$) stars, respectively.\\

\subsection{Dating the dSph infall from their orbital properties}

Within a 60 kpc radius region in the MW halo, dwarfs are the halo population with the most energetic orbits. They are also the only known population inhabiting regions well beyond this radius, which further supports a different infall time than that of other halo populations. Assuming most halo populations are coming from former infall, \citet{Boylan-Kolchin2013} showed from numerical simulations ({\it Aquarius}) that their infall time can be dated through their 3D velocity or their total energy. This is especially expected for material brought by infalling dwarfs or satellites, which are subjected to friction caused by, e.g., tidal effects. Even if {\it Aquarius} simulations do not include the gas, one may expect that ram pressure may additionally reduce the total energy of infalling gas-rich dwarfs. \citet[see their Fig. 2]{Hammer2020} used Gaia DR2 3D velocities of dwarfs to show that many of them were near 1.15 $\times$ $V_{\rm vir}$, a value consistent with a recent, less than 4 Gyr, infall according to \citet{Boylan-Kolchin2013}. However, this was based on a small number of only seven dwarf galaxies, with relatively large uncertainties, which prevented a firm conclusion.\\

\begin{figure}
\includegraphics[width=3.43in]{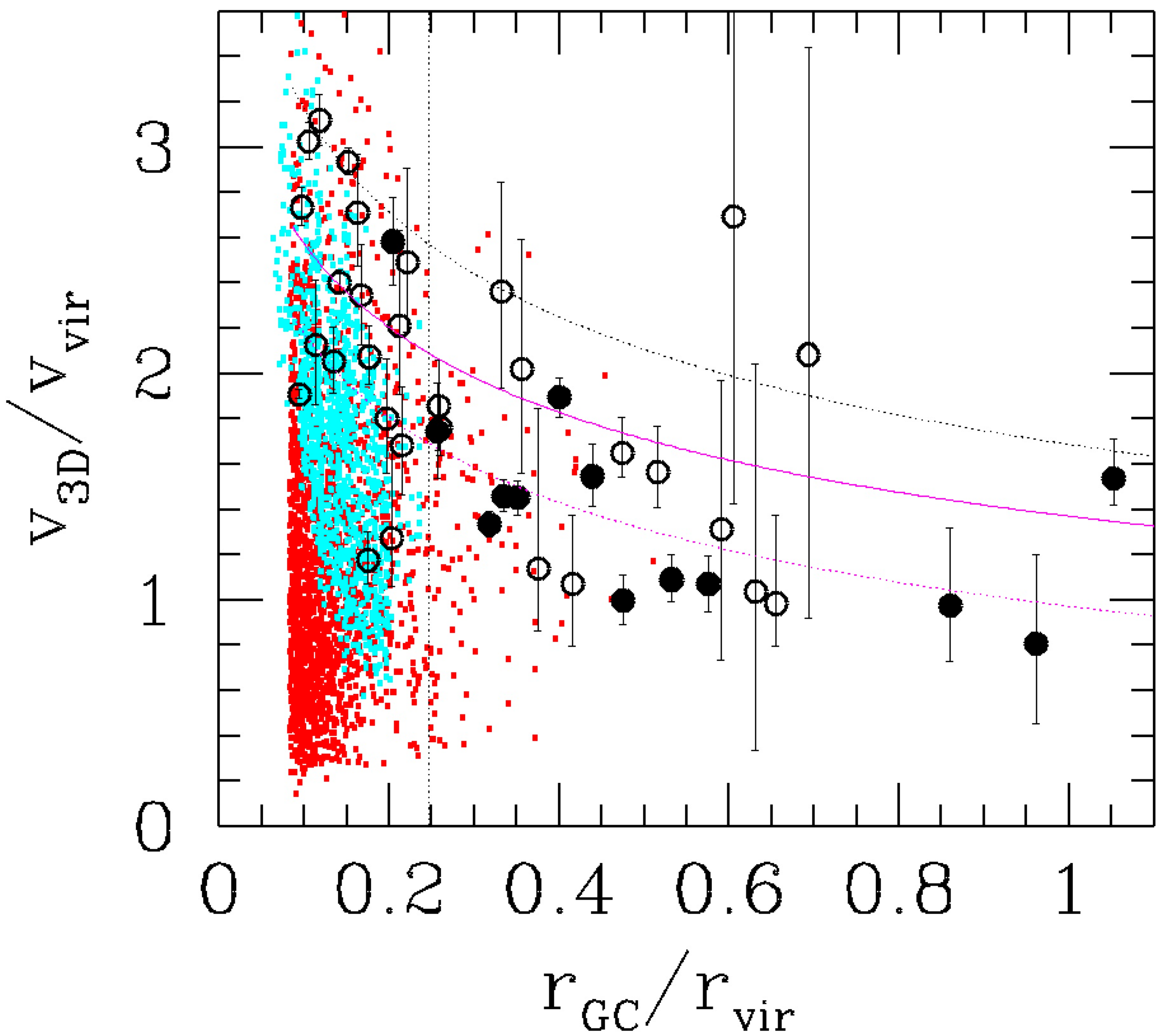}
\caption{Phase diagram of dwarfs (open and full black dots), K-giant stars (small red squares), and Sgr stream stars (small cyan squares). Axes are based on \citet{Boylan-Kolchin2013}'s framework, using the \citet{Bovy2015} MW mass model as in \citet[see their Figure 2]{Hammer2020}. Note that the \citet{Bovy2015} model is similar to that of \citet[see a comparison of the two mass profiles in Figure 1 of \citealt{Hammer2020}]{Eilers2019}, with an MW total mass of $8.1\times10^{11}$ $M_{\odot}$. Comparison with the later DR2 study shows the considerable increase of dwarfs of high velocity, especially at low radii. As in \citet{Boylan-Kolchin2013} the solid magenta line represents a curve of constant energy with $V_{\rm 3D}= 1.15 V_{\rm vir}$ and the dotted magenta curve delimits the recently,$<$ 4 Gyr, accreted subhalos as expected from the {\it Aquarius} simulation (see the text). The dotted black curve represents the escape velocity, $v_{esc}=\sqrt{-2\Phi}$. The dotted vertical line indicates $r_{\rm GC}/ r_{\rm vir}$= 0.245 (i.e., 60 kpc).}
\label{fig:BK}
\end{figure}

Let us now analyze with Gaia EDR3 the behavior of each population in the inner 60 kpc (or  $r_{GC}/ r_{vir}$= 0.245) region in Figure~\ref{fig:BK} (see the modeling details in the caption). It includes a much higher number (17) of dwarfs, with well reduced uncertainties. Twelve of them (70\%) are in between the two lines defining a recent infall. This compares to 22.5 and 12\% for Sgr stream and K-giant stars, respectively, meaning that the latter stars are more dynamically relaxed than are the MW dwarfs. K-giant stars likely include the Gaia-Sausage-Enceladus (GSE) stars corresponding to a former merger 8-12 Gyr ago, and this is confirmed by the corresponding small angular momentum found by \citet{Naidu2020}. There are several clues that Sagittarius infall occurred 3.5 to 5 Gyr ago\footnote{\citet{Bonifacio2004}  argued that a population at the age of $\sim$ 1 Gyr old is necessary to explain the solar metallicity population observed on the RGB of this galaxy and the blue plume observed in the color-magnitude diagram (see Figure 3 of \citealt{Bonifacio2004}). However, this can be consistent with the gas removal of a massive Sagittarius progenitor during its orbit within the MW halo.}, at an epoch it has formed more than half its stars before a complete shutdown \citep{Weisz2014}, and this time is also suggested by the most precise modeling \citep{Vasiliev2021b}.\\

The high energies and angular momenta of the dwarfs suggest that they have reached quite recently, $<$ 4 Gyr ago, the MW halo, from comparison to dark-matter cosmological simulations (see Figure~\ref{fig:BK}). Comparing them further to old ($>$ 8 Gyr, K-giant stars) and relatively recent (3.5-5 Gyr, Sagittarius stream stars) events, dwarfs appear to be the last newcomers into the MW halo. Their median total energy and angular momentum) is well above that of other halo populations (see Figure~\ref{fig:histos}), suggesting an average infall time significantly smaller than 3.5-5 Gyr. This can be accommodated with a first passage, knowing that for most dwarfs one orbit may take 1-2 Gyr. \\

We remark that all the above reasoning is based on the hierarchical scenario (though not forcedly the $\Lambda$CDM), i.e., assuming that dwarf galaxies inhabiting the halo of massive galaxies have progressively fallen in with time, which left an imprint on their angular momentum and energy properties.

\section{A possible structure related to the Sagittarius orbit?}

\begin{figure*}
\includegraphics[width=6.8in]{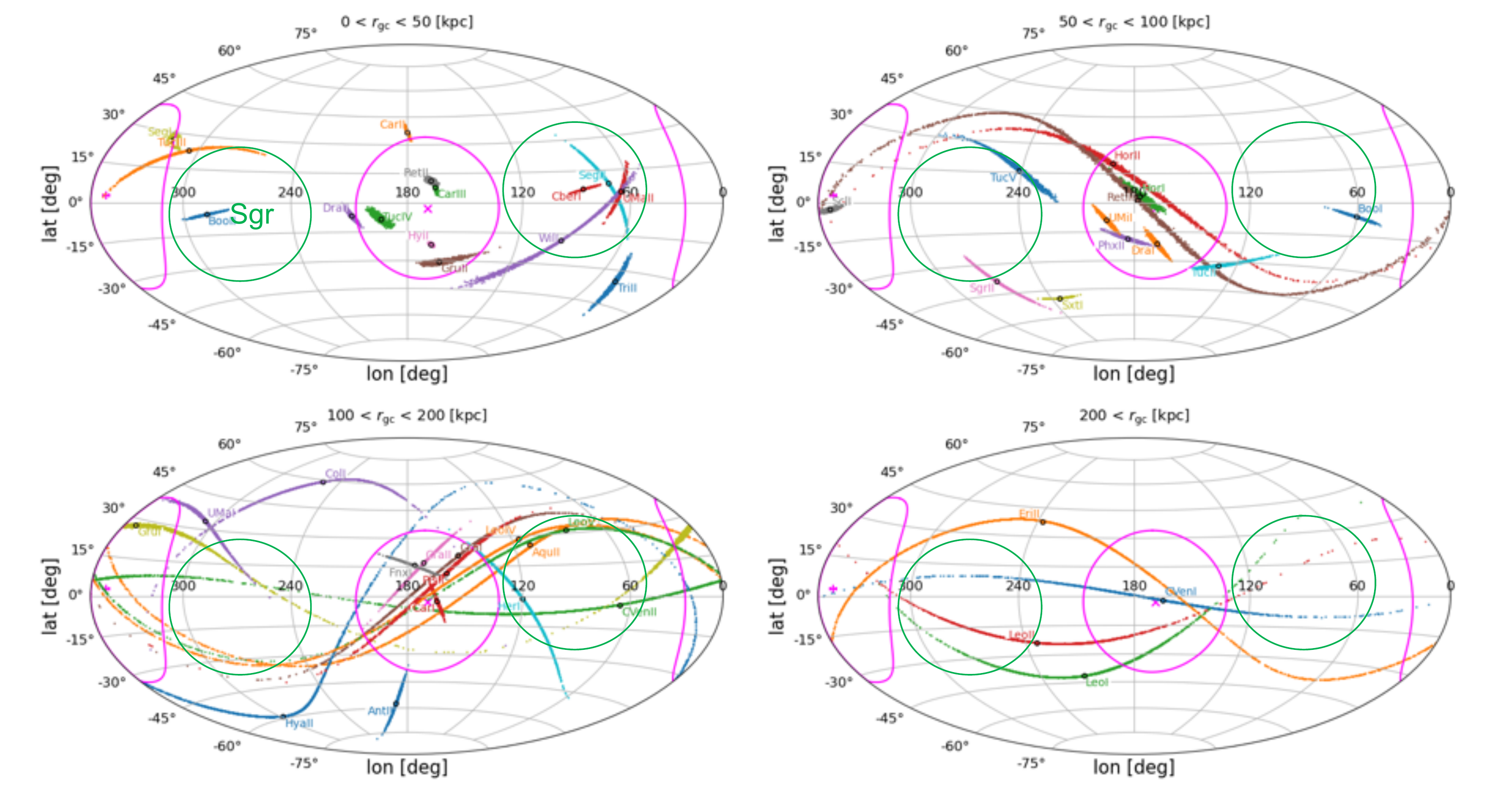}
\caption{ Angular momentum poles of dSph orbits in an Hammer-Aitoff sky projection (Galactocentric coordinates), using four distance intervals from the Galactic center (from left to right, top to bottom, 0$<$ $r_\mathrm{GC}$ $<$ 50 kpc, 50$<$ $r_\mathrm{GC}$ $<$ 100 kpc, 100$<$ $r_\mathrm{GC}$ $<$ 200 kpc, and 200$<$ $r_\mathrm{GC}$. The  magenta circle defines the VPOS location as in \citet{Li2021}, with magenta lines defining the opposite direction. Similar circles in green delineate the same area at the Sagittarius position, and at 180 degrees for selecting possible counter-rotating dwarfs.  Small points around each dSph plot the orbital poles from 2000 Monte Carlo simulations, for which the dot represents the median. }
\label{fig:VPOS_SAG}
\end{figure*}

Figure~\ref{fig:VPOS_SAG} is similar to Figure 2 of \citet{Li2021} providing accurate pole positions of MW dwarfs from EDR3 in an Hammer-Aitoff sky projection. It shows that besides a prominent VPOS, there are five dwarfs with orbital poles inverted from that of Sagittarius. They are Bootes I, Coma Berenices, Segue II, Ursa Major II, and Willman,. Moreover we find that Bootes II, Sagittarius II, Tucana III and Tucana V are orbiting along the same direction as Sagittarius. If confirmed, this would reveal an additional large-scale structure, including $\sim$ 20\% of the dwarfs, and almost perpendicular to both the VPOS and the MW disk. This Sagittarius polar structure (dubbed as 'SPOS') is, however, less prominent than the VPOS \citep{Pawlowski2014} that includes the majority of MW dwarfs, and most notably the most luminous ones.\\

\begin{figure}
\includegraphics[width=3.4in]{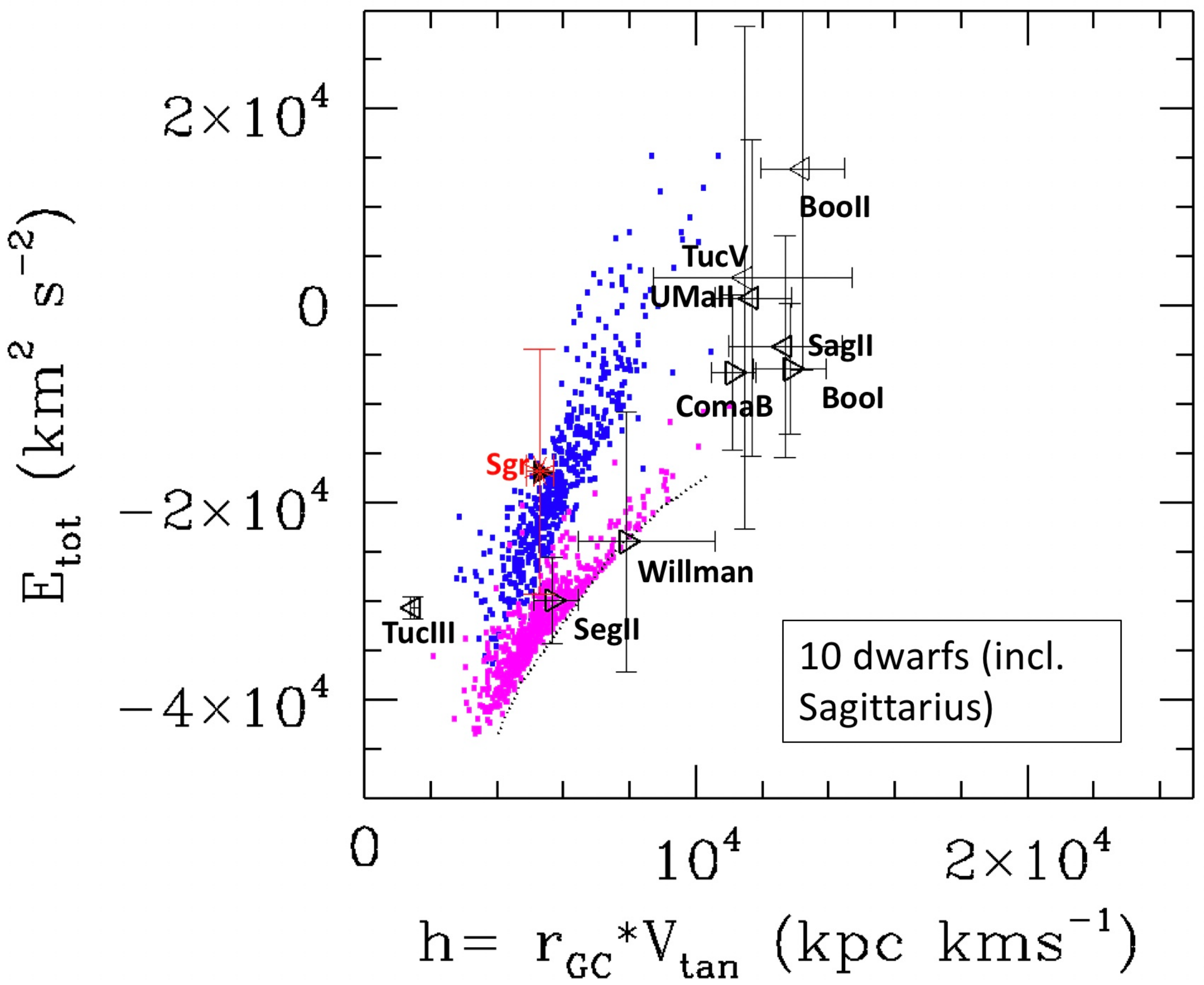}
\caption{ Location in the ($h$, $E_{\rm tot}$) plane of dwarfs with orbital poles aligned or counter-aligned with that of Sagittarius. Small magenta and blue points represent stars of the Sagittarius stream, distinguishing the Leading (Sagittarius longitude $\Lambda$ $>$ 12 degrees) and the Trailing Arm ($\Lambda$ $<$ -12 degrees), respectively.} 
\label{fig:E_h_SAG}
\end{figure}

Figure~\ref{fig:E_h_SAG} compares the location in the ($h$, $E_{\rm tot}$) plane of Sagittarius stream stars (magenta and blue dots) to that of dwarfs possibly associated with Sagittarius. Except Segue II and Tucana III (and Sagittarius), they have much larger energy and angular momentum than stream stars, which prevents them from being debris associated with the streams. This is true also for Segue II and Willman, because their $h$ and $E_{\rm tot}$ values place them very near the Leading Arm, while their projections on the sky are closer to the Trailing Arm. As for the VPOS \citep{Pawlowski2014}, the presence of counter-rotating dwarfs in the Sagittarius orbit implies that they spread out over a large circle passing on both sides of the MW center. However, it is less vast since it does not include dwarfs at distances larger than 65 kpc, so its nature could be more difficult to establish.  On the one hand, Coma Berenices, Bootes I, and perhaps Ursa Major II and Willman, seem to share almost a common orbit given their pericenters and apocenters as found in Table 4 of \citet{Li2021}. These galaxies are  counter-orbiting with Sagittarius, and could have been accreted as one group. On the other hand, Figure~\ref{fig:E_h_SAG} shows that in comparison to other members of this putative group, Willman has a small energy and angular momentum, and Ursa Major II has an inverted radial velocity.
More precise orbital measurements are necessary to verify whether the SPOS is restricted to dwarfs in the inner halo, and also to check if some of the associated dwarfs may have been members of a common group. Comparing Figure~\ref{fig:E_h_SAG} with Figure~\ref{fig:E_AM_allpop} indicates that the low energy-low angular momentum area of Figure~\ref{fig:E_AM_allpop} is mostly populated by SPOS galaxies. This is consistent with the fact that Sagittarius-associated galaxies are more dynamically relaxed because they came earlier. More statistics are needed to robustly conclude on this point.

\section{Discussion}
\subsection{MW dwarfs: the latest newcomers in the halo}
MW dwarf galaxies have much larger total energy and angular momentum than K-giants, globular clusters, and Sgr stream stars, a result obtained after having homogenized the number density function distribution. We have been very strict and  eliminated dwarfs with uncertainties in the ($h$, $E_{\rm tot}$) plane that are too large. Relaxing this criterion would only increase their $h$ and $E_{\rm tot}$ excesses, leading to smaller probabilities that their distributions could be consistent with that of MW halo stars and GCs. This is because dwarfs with large uncertainties have the bias to be seen as even more energetic and with higher angular momentum (see Appendix~\ref{sec:h_E_err}). We also remark that tangential velocities and their errors have been estimated through a full Monte Carlo estimation of the three velocity components ($v_r$, $v_{\theta}$, and $v_{\phi}$) that can take  positive and negative values (see Table 2 of \citealt{Li2021}). This means that their large values, as well as that of the angular momentum, cannot come from biases linked to estimates of positive quadratic quantities. \\

 In the hierarchical paradigm, the most energetic bound dwarfs are expected to be the latest newcomers \citep{Boylan-Kolchin2013}. Dwarfs are likely bound for MW total mass larger than 8 $10^{11}$$M_{\odot}$ (\citealt{Li2021} and see Figure~\ref{fig:phase}), which is also true for most of the classical dSphs for MW total mass larger than 5 $10^{11}$$M_{\odot}$ (see full triangles in Figures~\ref{fig:phase} and ~\ref{fig:E_AM_allpop}). As noticed by \citet{Boylan-Kolchin2013}, being bound does not exclude a first passage, i.e., a first infall needs not, and almost always does not, imply an unbound orbit. It suggests that the MW dwarfs have been more recently gravitationally attracted by the MW than any other halo populations. This includes stars in the stream associated with Sagittarius, the relatively recent infall of which would then suggest a first infall for the more energetic MW dwarfs. \\

\subsection{Could a late arrival of dwarfs be reconciled with their star formation histories?}
A possible argument against a first infall of dwarfs in the MW potential is the claim that they lack a young stellar population. The argument relies on the hypothesis that at the start of the infall the galaxy was gas rich \citep{Yang2014,Hammer2018a} and lost the gas due to ram pressure. The ram pressure is expected to produce a burst of star formation that consumes whatever gas is not stripped \citep[see, e.g.,][]{Kapferer2009}.
The applicability of both hypotheses can be questioned. A galaxy like ultra-faint dwarf galaxies, of very low mass, even evolving in isolation,  may well lose all its gas due to the galactic winds triggered by a short star formation burst. Furthermore, while the effect of ram pressure on star formation is well documented both theoretically and observationally for large gaseous galaxies, it is  unknown in the case of extremely low mass galaxies. \\

Let us concentrate on the claim that among the nine classical dSphs, four of them (Sculptor, Draco, Ursa Minor, and Sextans) comprise only populations of age larger than 8 Gyr. How robust is this claim? All the observed dwarf spheroidal galaxies have indeed a blue plume, which could be interpreted as evidence of a young population.  However, the most commonly accepted interpretation is that this blue plume, is due to blue straggler stars (BSS). 
There are two channels for the formation of a BSS: mass transfer (including coalescence) in a binary system or star-star collision. For field stars it is firmly established that the dominant mode of formation is from binary star evolution \citep{Preston2000}, while in GCs, the dense environment may favor the star-star collisions.\\

For dwarf galaxies  there are two comprehensive studies of these populations \citep{Momany2007,Santana2013}. Both studies agree that the BSS fractions in  dwarf galaxies are distinctly different from those in GCs of similar luminosity \citep{Piotto2004}. 
The dwarf galaxies always show a larger fraction of BSSs. However, while  \citet{Momany2007} find an anticorrelation of  BSS fraction with integrated magnitude, like for GCs \citep{Piotto2004} 
albeit with a  smaller slope, \citet{Santana2013} find a constant fraction of BSS, whatever the magnitude of the galaxy. Both \citet{Momany2007} and \citet{Santana2013} 
conclude that BSSs are formed only from binary evolution without a strong impact of star-star collisions.
\citet{Santana2013} claim that their results are not inconsistent with a
shallow decrease of the BSS fraction with increasing luminosity (mass) of the galaxy, 
similar to what was found by \citet{Momany2007}. Yet a physical explanation
of such a trend is not simple to find.
Stellar collisions that disrupt binary systems that could evolve to be
BSSs, like those invoked in  GCs, appear to be events that are too rare in the low density
environment of a dwarf galaxy to play any role.

We remark that in fact both a constant fraction of BSS counts (that to be objective we prefer to call  blue plume counts) or a low decrease with luminosity can be naturally accommodated if the stars are instead a young population. Either the young population is a constant fraction of the total galaxy population, or it decreases slightly with increasing galaxy mass. This can be understood in terms of the mass of residual gas available to form the young population.  More massive galaxies could trigger more powerful winds and lose larger fractions of gas. Either way the uncertainties in the modeling are large enough that both scenarios can be accommodated.
\citet{Santana2013} performed synthetic population simulations including both old and young populations, to constrain the properties of a young population that could form the blue plume. Their conclusion is that these populations should have an age of 2.5\,Gyr$\pm$ $0.5$\,Gyr and constitute $1\%-7\%$ of the total population of the galaxy. Such a small fraction is not unexpected if formed during a recent star formation event triggered by the ram pressure. Although  \citet{Santana2013} discard this possibility considering it unlikely, with hindsight, we can say that this scenario appears perfectly consistent with the interpretation that dwarfs are experiencing their first infall. 
A reexamination of the nature of the observed blue plumes is necessary to verify whether they can be consistent with being entirely or partially formed by a young stellar population. 

\begin{figure}
\includegraphics[width=3.4in]{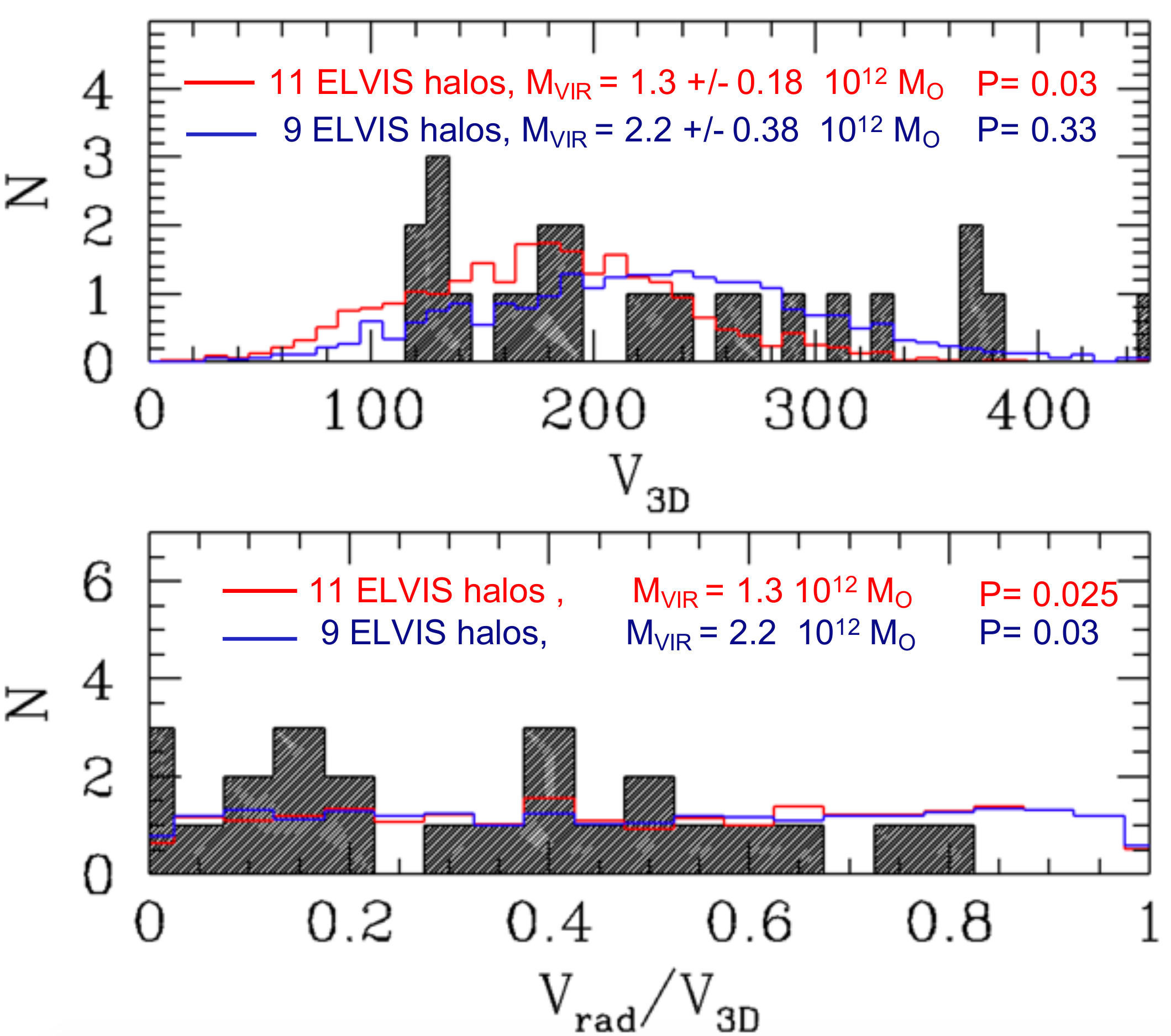}
\caption{Comparative histogram for  24 $r_{\rm GC} <$ 200 kpc and $L_v/L_{\odot}$ $>$ 4000 MW dwarfs (black shaded histogram) with 1116 and 1378 subhalos ($M_{vir}>$ 5 $10^{7} M_{\odot}$) lying within 200 kpc from the 11 lowest (red histogram) and the 9 highest (blue histogram) mass hosts of the {\it ELVIS} suite, respectively. Subhalo numbers have been rescaled to help the comparison. Kolmogorov-Smirnov probabilities that the simulated subhalos are coming from the same distribution as the observed one are given in the top left of each panel. Top: Distribution of dwarf 3D velocities (black shaded histogram) compared to that for subhalos of $M_{\rm vir}<$ 1.7 $10^{12} M_{\odot}$ hosts (red histogram), and of $M_{vir}>$ 1.7 $10^{12} M_{\odot}$ hosts (blue histogram). {\it Bottom:} distribution of $V_{\rm rad}/V_{\rm 3D}$ ratios for dwarfs (black shaded histograms), for subhalos of $M_{\rm vir}<$ 1.7 $10^{12} M_{\odot}$ hosts (red histograms), and and for subhalos of $M_{\rm vir}>$ 1.7 $10^{12} M_{\odot}$ hosts (cyan histogram). }
\label{fig:ELVIS}
\end{figure}

\subsection{Is there a dark-matter halo and subhalo system consistent with the MW and its dwarfs?}

We have shown in Section~\ref{sec:beta} that the tangential velocities of dwarfs are significantly in excess with $\beta$=-1.47$\pm$0.41 when compared to $\Lambda$CDM expectations for satellites ($\beta$= $0.25-0.45$, depending on the satellite faintness; see \citealt{Cautun2017}). Gaia EDR3 data are sufficiently precise to verify that this property is valid  both at small ($r_{\rm GC}<$ 100 kpc) and large distances ($r_{GC}>$ 100 kpc).  \citet{Cautun2017} estimates that these extreme values are expected in only $1.5\%$ of CDM satellites systems.  However, their sample was limited to few dwarfs, and in this paper, we have  revisited this question using the Gaia EDR3 results, confirming the tangential velocity excess for MW dwarfs at all distances.\\

We have considered the {\it ELVIS} suite of zoomed cosmological simulations \citep{Garrison-Kimmel2014}, which provides 10 pairs of host halos selected for their resemblance to the MW-M31 group. We have compared
the locations of sub-halos to that of MW dwarfs in the ($V_{\rm rad}$, $V_{\rm tan}$) plane. Following \citet{Rodriguez-Wimberly2021} we have accounted for the differences in distances between observed dwarfs and simulated subhalos, by limiting their distances to 200 kpc and their visible luminosity to be larger than 4000 $L_{\odot}$. The last criterion has the advantage of removing the ultra-faint dwarfs that can be difficult to detect at large distances, and a Kolmogorov-Smirnov test shows that simulated subhalos and observed dwarfs  are consistent with sharing the same distance distribution with $P= 0.33$. It results that we hardly find a distribution similar to that observed, although the discrepancy shrinks when the host halo mass increases. This prompted us to divide the  {\it ELVIS} sample accordingly to the host mass (see red and blue histograms in Figure~\ref{fig:ELVIS}), one with 11 host halos having $M_{\rm vir}<$ 1.7 $10^{12} M_{\odot}$ (average 1.3 $10^{12} M_{\odot}$), and the other with 9 host halos having $M_{\rm vir}>$ 1.7 $10^{12} M_{\odot}$ (average 2.2 $10^{12} M_{\odot}$).\\

The bottom panel of Figure~\ref{fig:ELVIS} shows that the relative contribution of the radial velocity to the 3D velocity does not depend on the host halo mass (compare red and blue histograms). It significantly differs from that observed for MW dwarfs, leading to Kolmogorov-Smirnov probabilities of 2-3\%. This confirms the estimate by \citet{Cautun2017} of the MW and its dwarfs' representativeness among systems of satellites in the CDM paradigm.  
Moreover, the observed distribution of dwarf 3D velocities is well reproduced by massive hosts with $M_{\rm vir}=$ 2.2$\pm$0.38 $10^{12} M_{\odot}$, and only poorly by lower mass hosts.  This would not come as a complete surprise, since mass estimates based on the assumption that dwarfs are satellites generally require high MW masses \citep{Boylan-Kolchin2013,Callingham2019}.\\

However, there is no known example for which such massive halos can fit the slightly declining MW rotation curve \citep{Eilers2019,Mroz2019}, and this result is coming from several studies having used different density profile or baryon contributions \citep{Eilers2019,deSalas2019,Karukes2020,Jiao2021}. In the  {\it ELVIS} suite, only the smallest  mass hosts of the {\it ELVIS} suite with $M_{\rm vir}=$ 1.3$\pm$0.18 $10^{12} M_{\odot}$ are consistent with the MW rotation curve. In other words, the MW dwarf 3D velocities are consistent with satellites of host galaxies at least twice as massive as the MW. Conversely, less massive hosts of the {\it ELVIS} suite with $M_{\rm vir}=$ 1.3$\pm$0.18 $10^{12} M_{\odot}$  if consistent with the MW rotation curve, possess satellites with 3D velocities inconsistent with the observed 3D velocities of dwarfs, as illustrated in the top panel of Figure~\ref{fig:ELVIS} (compare black shaded and red histograms, Kolmogorov-Smirnov test, with P= 0.03).\\ 

On the one hand, the only satellite systems that possess similar kinematics to the MW dwarfs are those of a host too massive to reproduce the MW rotation curve. On the other hand, for all {\it ELVIS} suites of zoomed cosmological simulations, simulated satellites have too large radial (or too small tangential) velocity contribution when compared to that of MW dwarfs. Because the satellite physics is only controlled by the gravitational force, this suggests that another mechanism is acting to differentiate MW dwarfs from satellite subhalos.




\begin{figure}
\includegraphics[width=3.4in]{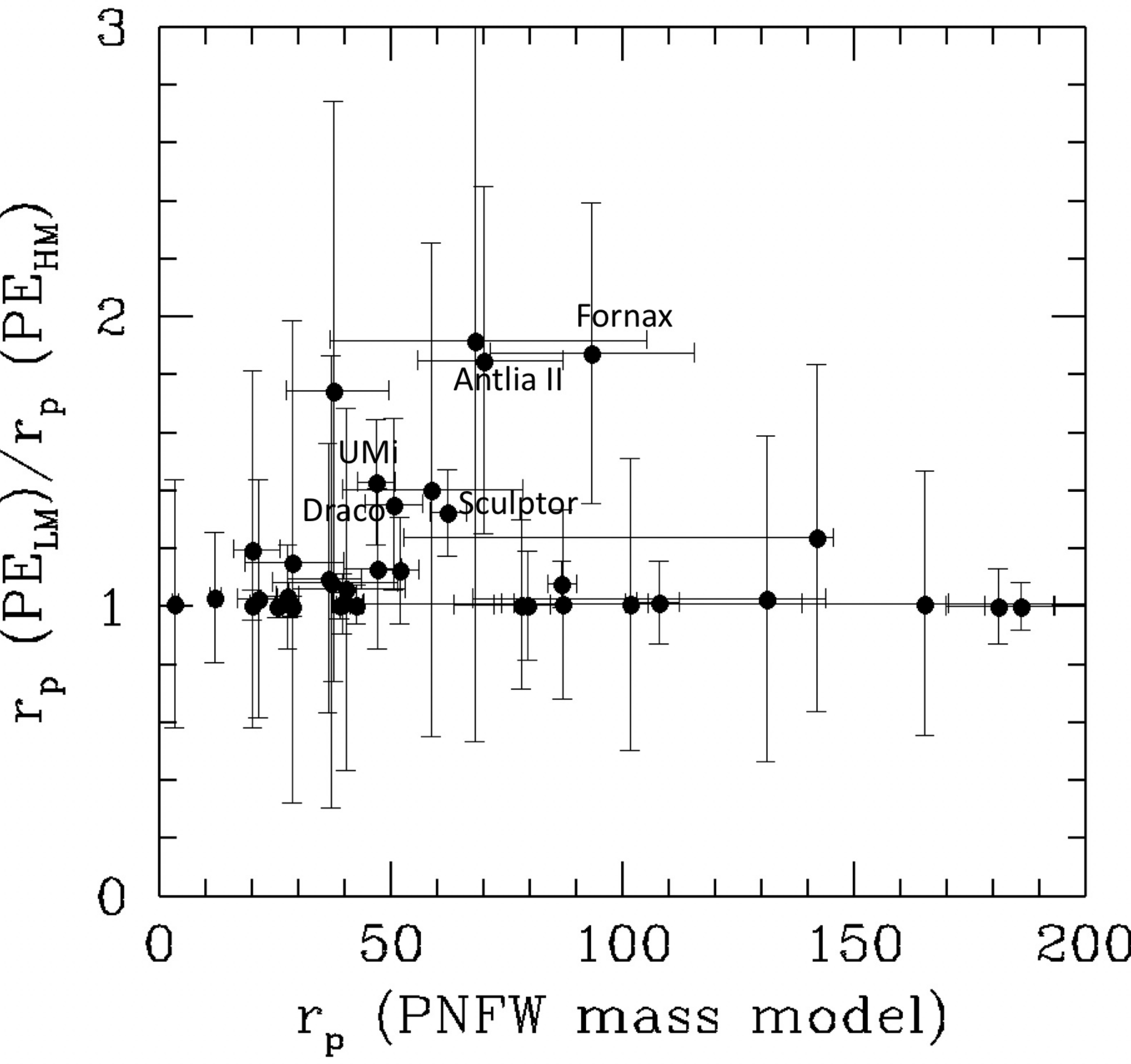}
\caption{Ratio of pericenters derived for the lowest mass model of the MW, PE$_\mathrm{LM}$ ($M_{\rm tot}$= $2.8\times10^{11}$ $M_{\odot}$), divided by that derived for the highest mass available to fit the MW rotation curve \citep{Jiao2021}, PE$_\mathrm{HM}$($M_{\rm tot}$= $15\times10^{11}$ $M_{\odot}$). This ratio is plotted against the pericenter derived for the intermediate mass model proposed by \citet{Eilers2019}. The few objects showing changes of their pericenters with the MW mass are labeled, while nine dwarfs with $\delta$$r_{p}$ $>$ $r_{p}$ are excluded for a better view, namely Canes Venaciti I  and II, Grus I, Leo I, Leo II, Leo IV, Phoenix II, Reticulum III, and Willman I. }
\label{fig:rp}
\end{figure}

\subsection{Could Gaia EDR3 orbital motions have been impacted by ram-pressure effects?}
Dwarf angular momentum and total energy may vary due to energy exchanges with the MW, e.g., through tides and ram pressure if dwarf progenitors were gas-rich\footnote{The orbit of much more massive dwarfs such as the LMC could be further decayed by the back reaction of the host, through gravitational torques \citep{Weinberg1986,Weinberg1989,Tamfal2021}}. Let us first concentrate on the first mechanism and on the nine classical dSphs for which observations are sufficiently deep  to test their outskirt properties. If dwarfs were MW satellites, their orbital energy would decay proportionally to their mass ratio to the MW mass \citep[see their Eq. 8.17]{Binney2011}. For a satellite, the orbital decay is expected to be accompanied by tidal stripping, and then by the formation of tidal tails, such as in Sagittarius. Conversely to that, very few other classical dSphs, and possibly none\footnote{Only Carina has been suggested with tides  \citep{Battaglia2012}, though contamination by LMC debris may discard it \citep{McMonigal2014}}, are known to be affected by tidal stripping, which can be then excluded as a major source of energy exchange between the MW and dwarfs.\\

 The bottom panel of Figure~\ref{fig:ELVIS} shows that for dwarfs, radial velocities are contributing less to their kinetic energy when compared to simulated subhalos.  If dwarf progenitors were gas-rich, ram pressure would first have slowed down their motions, and then remove their gas. For MW dwarfs, ram pressure has been so efficient that all the gas has been removed leaving gas-free objects. Removing most of the baryonic mass in a dwarf is a very complex process that affects most of its structure through hydrodynamical turbulence and shocks \citep{Roediger2008}. It may also affect its orbital motion, first by slowing its total velocity ($V_{\rm 3D}$), and second perhaps by affecting its radial more than its tangential motion. Intuitively, this is because the hydrodynamic effects seem more efficient in the inner part of the MW gas halo, potentially affecting the radial component more. Furthermore, by slowing down the dwarf motions, ram pressure may have circularized their orbits, reducing the radial component. \\
We have tested the above by using the \citet[see their Model 27 and MW2]{Wang2019}'s MW mass model including a halo gas component, which has been elaborated to successfully reproduce the Magellanic Stream. In the absence of ram pressure both tangential and radial velocities are expected to increase during the infall, due to the angular momentum conservation and to the MW gravity, respectively. We have simulated a single, DM-free SMC infalling into an orbit sufficiently tangential to allow it to be affected by ram pressure. We have calculated the velocity components after passage to pericenter (65kpc), when the SMC is lying at 180 kpc and has lost most of its gas.  Comparing the cases with and without the ram pressure effect, we find a decrease of $V_{\rm 3D}$, $V_{\rm rad}$, and $V_{\rm tan}$ by 33\%, 40\%, and 17\%, respectively.  \\

 
Although a full simulation study is beyond the scope of this paper, the above suggests that ram pressure could have reduced the contribution of radial velocities to the kinetic energy. If correct, this would have affected gas-rich  dwarf progenitors since their first entrance into the MW halo, and it can explain why the radial velocity contribution to the kinetic energy  differs from that of purely gravitationally dominated subhalos of the {\it ELVIS} suite (see the bottom panel of Figure~\ref{fig:ELVIS}). 
It is also possible that a reduction of radial velocities could help explain why many dwarfs lie near their pericenter \citep{Fritz2018,Simon2018}. In fact, at such a location $V_{\rm rad}$ is null, and a bias against radial velocities may help many dwarfs to reach their pericenters more easily. The proximity to pericenter is unexpected if dwarfs were long-lived MW satellites, a occurrence that is associated with a very small probability, $P$ $\sim$ $10^{-4}$ \citep{Hammer2020}. \citet{Li2021} recently built volume-complete samples, showing that this property cannot be explained by either a very large MW mass, or by statistical biases. Proximity of dwarfs to their pericenters could also explain why pericenter estimates are so independent of the MW mass models (see Figure~\ref{fig:rp} and also \citealt{Simon2018}). In fact, all dwarfs have their  pericenter changing by less than 10\% for the whole MW mass range made available by \citet{Li2021}, except Antlia II and Fornax (factor 2), and  Ursa Minor, Draco \& Sculptor (factor 1.3). \\

  
If ram pressure has slowed the dwarf progenitors, it implies that their initial energy was even higher before they entered the MW halo. This suggests an alternative scenario for which dwarf progenitors were on fast encounter orbits \citep{Binney2011} with the MW. After gas removal and due to the lack of gravity caused by the gas loss, stars likely expand following a spherical geometry. Fast encounter galaxies often obey to the impulse and distant approximations \citep{Aguilar1985}, and the orbital energy decay is modest because it is affected by second-order effects (heating or tidal shock term). This is because expanding stars on a spherical geometry allows first-order effects or diffusion term to vanish \citep{Binney2011}, which explains why tidal stripping is not observed. In such a case, dwarf internal motions would be dominated by tidal shocks, which mostly affect only the (fraction of) stars that are in resonance with the MW gravitational potential \citep{Weinberg1994}. In this scenario, dwarf progenitors have recently lost their gas and the observed dwarf internal kinematics\footnote{The whole mechanism is described in a video based on hydrodynamical, N-body simulations optimizing the dynamics of stars, see https://www.youtube.com/watch?v=m55iBXISYyE} are fully  explained by tidal shocks \citep{Hammer2019,Hammer2020}, without dark-matter contribution. This warrants that their internal crossing time is smaller than their encountering time with the MW (see a full description in \citealt{Hammer2019,Hammer2020}), which is a prerequisite for being in fast encounter conditions. 



\section{Conclusion}
We have examined the orbital behavior of MW dwarfs, including their total energy and angular momentum, both quantities that are expected to be conserved if there was no exchange of energy with their host, the MW. Since the angular momentum is independent of the adopted MW mass profile, and because we have investigated the whole range of MW profiles to reproduce its rotation curve, our results can be considered valid independently of the exact MW mass. We find that:
\begin{itemize}
\item Dwarf tangential velocities are significantly in excess ($\beta$= -1.47$\pm$0.41) when compared to $\Lambda$CDM predictions for satellites ($\beta$= 0.25-0.45), and this is the case for all distance ranges.
\item Having compared to two suites of cosmological simulations, we have found no combination of dark-matter host halo and associated subhalos from cosmological simulations able to reproduce both the MW dwarfs with high kinetic energy and the MW rotation curve.
\item At small distances ($<$ 60 kpc), dwarfs have unexpectedly large energies and angular momenta when compared to expectations for long-lived satellites, pointing toward infall times smaller than 4 Gyr.
\item Their energies and angular momenta are significantly larger than that of stars coming from the relatively recent infall of Sagittarius, $3.5-5$ Gyr ago.
\end{itemize}

Orbital motions from Gaia EDR3 can be considered as very robust for a significant number of dwarfs (40 in total, 17 within 60 kpc). The resulting analysis supersedes other indirect clues such as the star formation history of  few classical dwarfs, which from their old stellar age distribution have supported a scenario in which dwarfs are long-lived satellites of the MW. On the contrary, we conclude that due to their unequaled high energies and angular momenta, most dwarfs cannot be long-lived satellites, and if they could be bound to the MW, they are at first passage, i.e., infalling less than $\sim$ 2 Gyr ago. We also suggest a reevaluation of the dwarf SFHs, in particular of the nature of their blue plume.\\

Gaia EDR3 velocities are sufficiently precise to allow a full comparison of their velocities to modeling, indicating for the first time that gravitational forces are not sufficient for explaining their orbits. We suggest that ram pressure could be the additional mechanism that explains their deficit of radial velocities when compared to satellite systems only dominated by gravity. It may also help us to understand why dwarfs lie near their pericenters, where radial velocities are small. This could lead to scenarios in which they would be fast encounters to the MW, for which their intrinsic high kinematics is caused by tidal shocks instead of dark matter.\\

In addition, we have possibly identified the presence of a vast structure in which dwarfs are orbiting or counter-orbiting with the Sagittarius dSph that we have called SPOS, which includes 20\% of the MW dwarfs. More data and better precision are needed to verify whether it could be a structure similar to the VPOS, and to compare robustly whether or not their orbital properties may differ from that of the majority of dwarfs in the VPOS. \\

Many important questions about MW dwarfs  remain open, including why they lie in vast polar structure(s). Given the difficulty of interpreting this question in the $\Lambda$CDM context, other models could have to be considered. Alternatively, the VPOS could be related to the suggestion \citep{Kroupa2005,Metz2007,Pawlowski2012} that dwarf progenitors could belong to tidal tails of former major interactions between massive galaxies. This is because tidal tails are coherent vast structures sharing a common orbital plane \citep{Pawlowski2018} making it possible to form vast polar structures such as the VPOS. It could be interesting to further examine whether or not, during the last $\sim$ $1-2$ Gyr, the MW could have been affected by the nearby passage of one such structure \citep{Fouquet2012,Hammer2013,Yang2014}, possibly related to the M31 disk rejuvenation after its recent merger \citep{Hammer2005,Hammer2018b}.

\begin{acknowledgments}
We are very grateful to the referee whose remarks have helped to finalize the structure of the paper and of its arguments. 
\par This work has made use of data from the European Space Agency (ESA) mission  {\it Gaia} (\url{https://www.cosmos.esa.int/gaia}), processed by the {\it Gaia} Data Processing and Analysis Consortium (DPAC,
\url{https://www.cosmos.esa.int/web/gaia/dpac/consortium}).
\par Funding for the DPAC has been provided by national institutions, in particular the institutions participating in the {\it Gaia} Multilateral Agreement. We are thankful for the support of the International Research Program Tianguan, which is an agreement between the CNRS, NAOC and the Yunnan University.  M.S.P. thanks the Klaus Tschira Stiftung gGmbH and German Scholars Organization e.V. for support via a Klaus Tschira Boost Fund. 
\end{acknowledgments}

\appendix

\section{Variations of properties with the MW mass}
\label{sec:MWmasses}

\begin{figure*}
\includegraphics[width=6.0in]{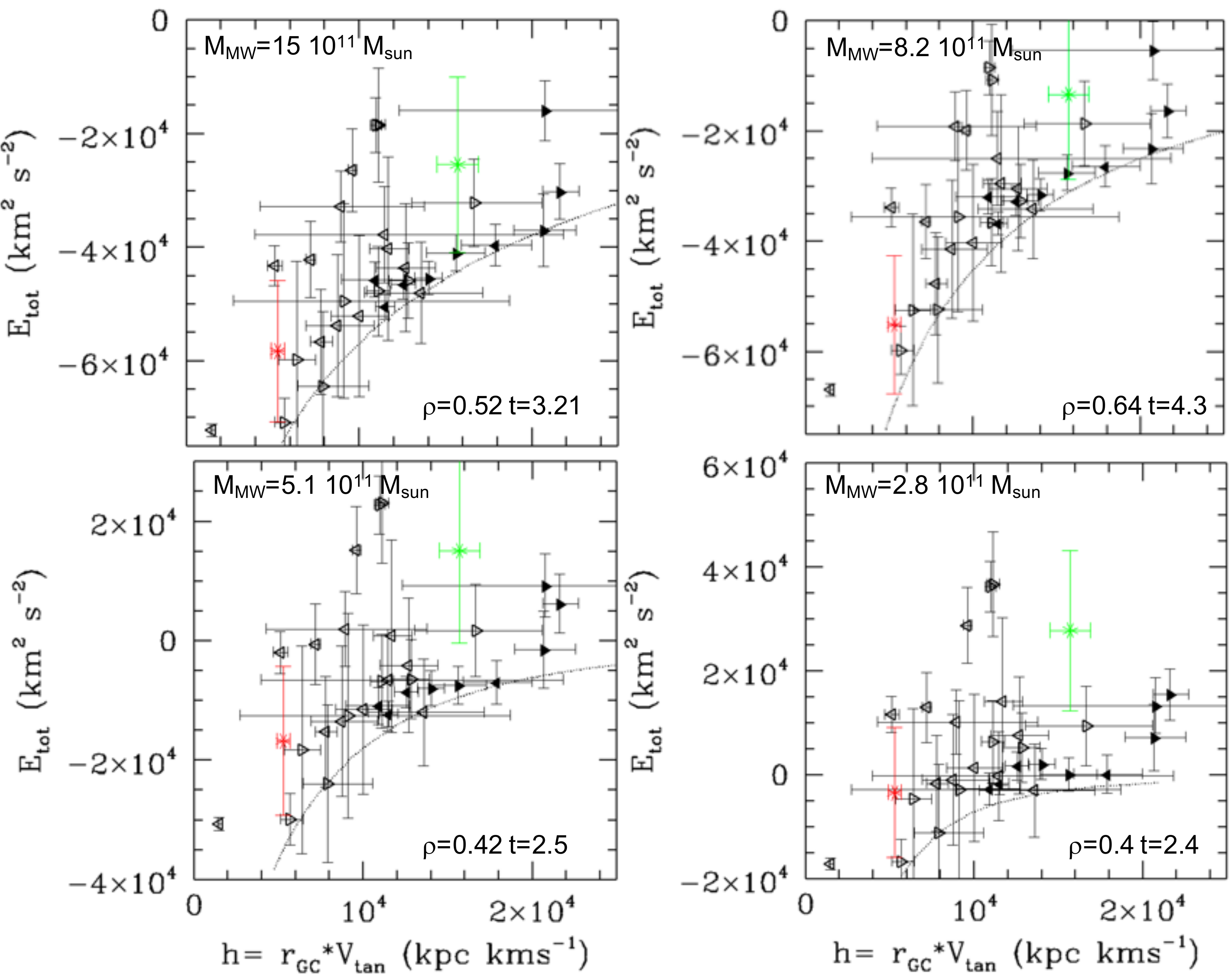}
\caption{Total energy vs. angular momentum for the 30 dwarfs in Figure~\ref{fig:E_AM_allpop}, and assuming the four different Milky Way mass models (see \citealt{Li2021}). dwarfs are shown by triangles as in Figure~\ref{fig:phase}.  Panels from left to right, top to bottom: decreasing total MW mass indicated in the top left of each panel, as well as the significance of the correlation (and t parameter) on the bottom right.
}
\label{fig:E_AM_MWmass}
\end{figure*}

 Figure~\ref{fig:E_AM_MWmass} presents the dSph distribution for the four MW mass profile investigated by \citet{Li2021}. In the figure, only dwarfs with error of $h$ smaller than $10^{4}$ kpc $\times$ km $s^{-1}$ (error smaller than 2 $10^{4}$ kpc $\times$ km $s^{-1}$ on energy) are shown. Higher MW masses lead to larger positive correlations,  which is not unexpected given their mutual positive dependency on $r_{\rm GC}$. Figure~\ref{fig:E_AM_MWmass} also shows that many dwarfs seem to be on a line that approximately relates Tucana III to Sextans, with a slope increasing with the MW mass. \\

\section{Dependency of dSph distribution on the uncertainties}
\label{sec:h_E_err}
\begin{figure*}
\includegraphics[width=6.0in]{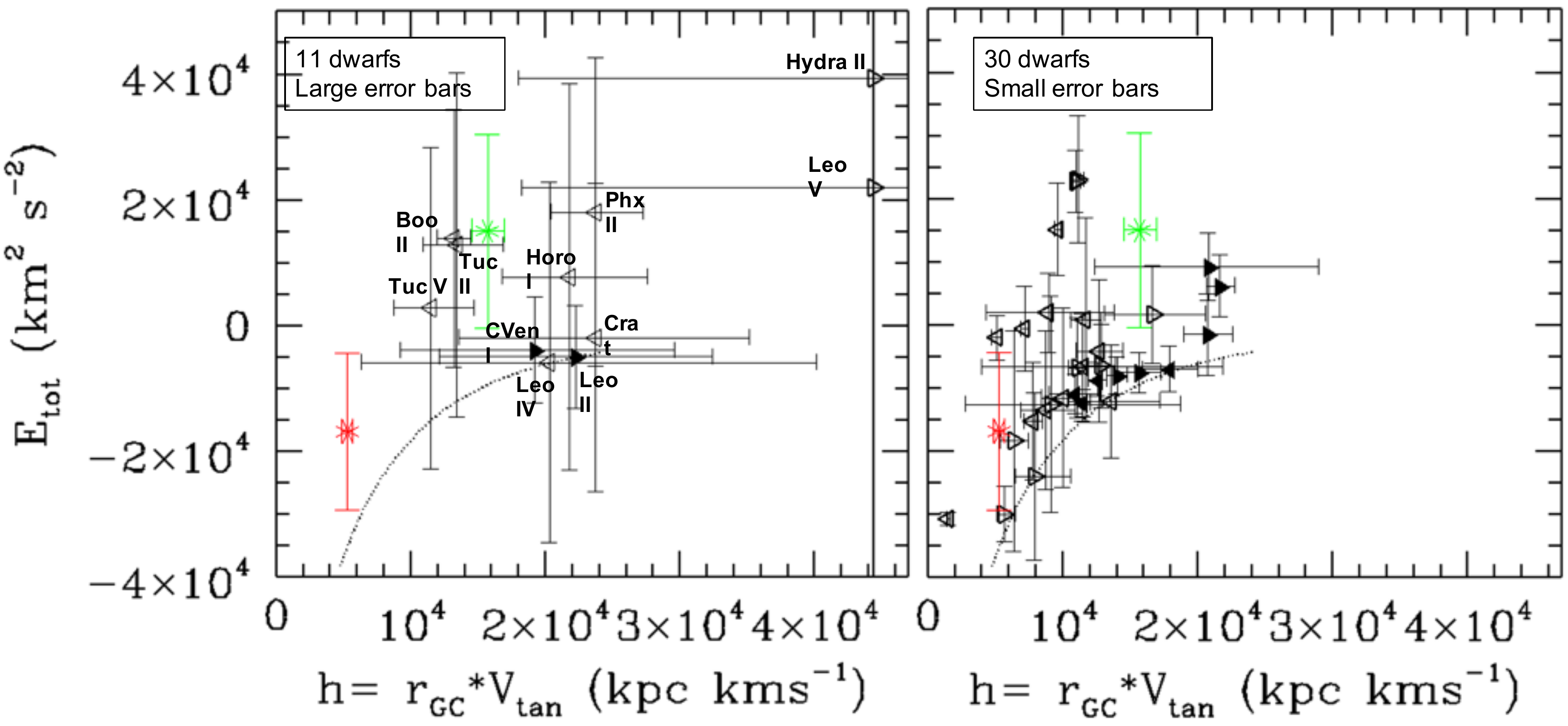}
\caption{Total energy vs. angular momentum for 11 (left panel, with labels) and 30 (right panel) dwarfs, with large and small uncertainties, respectively. Limits in uncertainties are $\Delta h$ $=$ $10^{4}$ kpc $\times$ km $s^{-1}$ and $\Delta E_{tot}$ $=$ 2 $10^{4}$ kpc $\times$ km $s^{-1}$, as defined in Figure~\ref{fig:E_AM_allpop}. The figure assumes the Milky Way intermediate mass model of  \citet[see also Wang et al. 2021]{Li2021}, and dwarfs are shown by triangles as in Figure~\ref{fig:phase}).
}
\label{fig:E_AM_errors}
\end{figure*}

Figure~\ref{fig:E_AM_errors} compares the locations in the energy-angular momentum plane of dwarfs with large (left) and small (right) uncertainties.   

\bibliography{mybib}{}
\bibliographystyle{aasjournal}



\end{document}